\documentclass[journal=jpccck,manuscript=article]{achemso}

\usepackage{graphicx}
\usepackage{natbib}
\usepackage{chemformula} 
\usepackage[T1]{fontenc} 

\usepackage{csvsimple}

\usepackage[version=3]{mhchem}
\usepackage{amsmath}
\usepackage{amssymb}

\usepackage{rotating}
\usepackage{subcaption}

\usepackage{numprint}
\usepackage{color}

\usepackage[final]{changes}

\makeatletter
	\newcounter{reaction}
	\renewcommand\thereaction{r\arabic{reaction}}
	\newcommand\reactiontag%
	{\refstepcounter{reaction}\tag{\thereaction}}
	\newcommand\reaction@[2][]%
	{\begin{equation}\ce{#2}%
	\ifx\@empty#1\@empty\else\label{#1}\fi%
	\reactiontag\end{equation}}
	\newcommand\reaction@nonumber[1]%
	{\begin{equation*}\ce{#1}\end{equation*}}
	\newcommand\reaction%
	{\@ifstar{\reaction@nonumber}{\reaction@}}
\makeatother

\author{Malte D\"{o}ntgen${}^\dagger$}
\affiliation{${}^\dagger$Chair of High Pressure Gas Dynamics, Shock Wave Laboratory, RWTH Aachen University, 52056 Aachen, Germany}
\phone{+49 241 80 26907}
\email{doentgen@hgd.rwth-aachen.com}

\author{K. Alexander Heufer${}^\dagger$}
\affiliation{${}^\dagger$Chair of High Pressure Gas Dynamics, Shock Wave Laboratory, RWTH Aachen University, 52056 Aachen, Germany}

\title[]
  {Non-Boltzmann Heat Transfer Between a Monoatomic Gas and a Solid Nanostructure}

\begin{document}

\begin{tocentry}
\includegraphics[width=3.25in, height=1.75in]{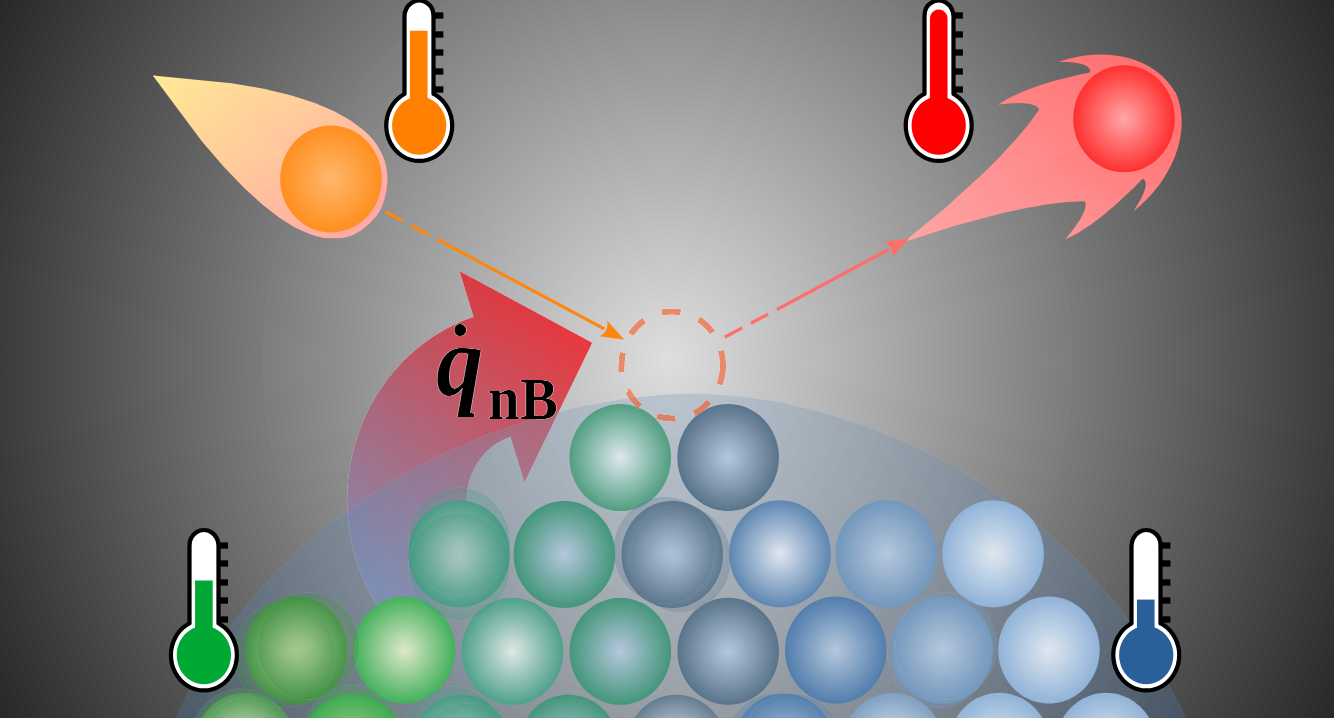}
\end{tocentry}

\abstract{
The effect of non-Boltzmann energy distributions on the pressure, impingement rate, and heat flux of a monoatomic gas in contact with a solid surface is investigated via theory and simulation.
First, microcanonical formulations of the pressure, impingement rate, and heat flux are derived from first principles and integrated with prototypical energy distributions.
Second, atomistic molecular dynamics simulations of an iron nanowire in a low-pressure argon atmosphere are used to test the non-Boltzmann heat flux theory.
While pressure is found to be unaffected by the energy distribution of the gas, the impingement rate increases by up to 8.5\;\% in the non-Boltzmann case.
Most intriguing, non-Boltzmann energy distributions can lead to a negative heat flux, meaning that heat flows from the cold solid to the hot gas.
This non-Boltzmann heat flux effect is validated via the molecular dynamics simulations and the solid is found to be 46\;\% colder than the gas in case of an hypothetical equilibrium for the upper limiting non-Boltzmann energy distributions.
The present fundamental findings provide novel insights into the properties of non-Boltzmann gases and improve the understanding of non-equilibrium dynamics.
}

\section{Introduction}
Non-equilibrium energy distributions of gases are commonly induced in gas dynamical processes involved in atmospheric (re-)entry~\cite{Boyce1996, Kosareva2021} and hypersonic cruise~\cite{Doroshenko1992}, but also through the expanding flow within a regular nozzle~\cite{Louviot2015, Dudas2020}.
From a thermal engineering point of view, knowing the heat loads acting on the solid structures is critical for reliably designing the respective structures.
How these heat loads are affected by non-equilibrium, i.e. non-Boltzmann, energy distributions is not fully understood.
Here, the non-Boltzmann heat flux of a monoatomic gas to a solid surface is derived from first principles via microcanonical integration.
Atomistic molecular dynamics simulations of a nanostructure in Boltzmann and non-Boltzmann gases are utilized to validate the theoretical findings.
Although deeply counter-intuitive, the presented theory and simulations suggest that non-Boltzmann heat fluxes might be negative under certain conditions, meaning that heat flows from the cold solid to the hot gas.

So far, the only setup for which negative heat fluxes have been observed experimentally are quantum entangled systems to the best of the authors' knowledge.
In 2008, Partovi~\cite{Partovi2008} theoretically constructed two quantum entangled, i.e. correlated, macroscopic systems with different initial equilibrium temperatures and found that heat would flow from the cold to the hot system due to the correlation.
Partly building on Partovi's theory, Jennings and Rudolph~\cite{Jennings2010} theoretically investigated the role of correlations in entropy-decreasing events.
Recently, Bera~et~al.~\cite{Bera2017} formulated generalized laws of thermodynamics for correlated systems and assumed that correlations store work potential, which in turn contribute to the negative heat flux processes.
The theory of Partovi~\cite{Partovi2008} has recently been proven experimentally by Micadei~et~al.~\cite{Micadei2019}.
The authors generated two correlated quantum systems with initially different effective equilibrium temperatures via Nuclear Magnetic Resonance techniques and found that the colder system in fact transfers heat to the hotter system.
The authors explained the heat flux from cold to hot through the decrease of the mutual information of the correlated system.

In the field of gas dynamics, quantum correlation is typically not considered and no effects which might cause negative heat fluxes are known so far to the best of the authors' knowledge.
Recently, D\"{o}ntgen~\cite{Doentgen2023} investigated the effect of non-Boltzmann energy distributions on the heat capacity ratio of a monoatomic gas.
The author found that non-Boltzmann energy distributions can significantly affect the behavior of monoatomic gases.
Non-Boltzmann effects have recently been investigated intensely in the context of chemical kinetics~\cite{Labbe2016, Labbe2017, Doentgen2017, Doentgen2017b}, initiated by Klippenstein and co-workers~\cite{Burke2015b, Goldsmith2015a}.
Building on these studies, Labbe~et~al.~\cite{Labbe2016, Labbe2017} and D\"{o}ntgen~et~al.~\cite{Doentgen2017, Doentgen2017b} contributed to the fundamental modeling of non-Boltzmann chemical kinetics.

For the continuous description of energetic non-equilibrium states of monoatomic gases, the Boltzmann equation~\cite{Huang1987} is typically solved numerically, e.g. via Direct Simulation Monte Carlo (DSMC)~\cite{Bird1994}.
The Boltzmann equation, however, is typically limited to rarefied gas conditions~\cite{Cercignani1985}.
For dense conditions and interactions between different phases, however, solutions of the Boltzmann equation have not been well developed~\cite{Maruyama2000}.

Molecular dynamics simulations are widely used for gaining insights into heat transfer processes~\cite{Maruyama2000}, especially at nanoscales~\cite{Frijns2004, Park2009, Yang2022, Ye2022}.
Zhakhovsky~et~al.~\cite{Zhakhovsky2018} used molecular dynamics simulations to validate and enhance their numerical solutions of the Boltzmann equation for heat and mass transfer through evaporation and condensation at surfaces.
The authors used molecular dynamics in particular to model the dynamics at the fluid-solid boundary.
Frijns~et~al.~\cite{Frijns2004} investigated the heat transfer of a dense gas to micro- and nano-channels via combined molecular dynamics and DSMC techniques.
The authors reported that both simulation techniques yield the same results for the investigated nano-channel setup.
Recently, Yang~et~al.~\cite{Yang2022} investigated the collisional heat transfer between gold and copper nanoparticles and either gaseous \ce{He}, \ce{Ar}, \ce{H2}, or \ce{N2}.
The authors simulated differently sized nanoparticles and determined thermal accommodation coefficients consistent with previous theoretical and experimental work.

Building on the prior theoretical and simulation work~\cite{Doentgen2023}, the non-Boltzmann effects on pressure, impingement rate, and heat flux are investigated in the present work.
First, these properties will be derived for a non-Boltzmann energy-distributed monoatomic gas from first principles via the corresponding microcanonical formulations.
Second, the present formulations will be tested against well-established formulations for Boltzmann energy-distributed monoatomic gas.
Third, the theoretical non-Boltzmann heat flux will be validated via molecular dynamics simulations of a nanostructure in Boltzmann and non-Boltzmann gases.
It will be shown how non-Boltzmann energy distributions induce heat flux from the nominally colder solid nanostructure to the nominally hotter gas at low densities.
The present work reports a negative heat flux for a classical, i.e. non-quantum, system for the first time to the best of the authors' knowledge.

\section{Theory}
The collisional interaction of a monoatomic ideal gas with an arbitrary energy distribution with an ideal, continuous solid surface is described theoretically here.
Analogous to the recent work of D\"{o}ntgen~\cite{Doentgen2023}, the present work considers the Boltzmann energy distribution (B) as reference for systems in thermal equilibrium and the delta energy distribution ($\delta$) as upper limiting case of non-Boltzmann energy distributions.
The delta energy distribution has very little physical relevance, since any collisional interaction would disrupt this distribution.
However, it allows to quantify the upper limiting effect of non-Boltzmann energy distributions on collisional interactions.
The Gaussian energy distribution (G) will be used to represent non-Boltzmann energy distributions in between the two limiting cases.
All energy distributions are enforced to have the same total energy to ensure comparability between the different distributions.
Note, however, that the entropy converges to zero when shifting the energy distribution from the Boltzmann to the delta case~\cite{Doentgen2023}.
The effect of non-Boltzmann energy distributions on the collisional momentum transfer, i.e. pressure, the impingement rate, and the heat flux from the gas to the solid surface is investigated theoretically through integration of the respective microcanonical formulations with the different energy distributions.

\subsection{Pressure}
The kinetic theory of gases defines the pressure $P$ acting on a surface area $A$ as the momentum $\Delta p$ transferred through collision in direction $\vec{u}$ normal to the surface over time $\Delta t$.
This momentum is the sum of the inbound and outbound momenta of the $N$ colliding particles $\Delta p = \langle p_{\text{in}, \vec{u}} \rangle + \langle p_{\text{out}, \vec{u}} \rangle$, respectively.
These particles of mass $m$ traverse the distance $L$ towards the surface and back again in time $\Delta t = m \cdot L / \langle p_{\text{in}, \vec{u}} \rangle + m \cdot L / \langle p_{\text{out}, \vec{u}} \rangle$, respectively.
When assuming that the total averaged squared momentum $\langle p_i^2 \rangle = \langle p_{i, \vec{u}}^2 \rangle + \langle p_{i, \vec{v}}^2 \rangle + \langle p_{i, \vec{w}}^2 \rangle$ is equally distributed over all three orthogonal spatial directions $\vec{u}$, $\vec{v}$, and $\vec{w}$, the averaged squared momentum in each direction is given by $\langle p_{i, \vec{u}}^2 \rangle = \langle p_{i, \vec{v}}^2 \rangle = \langle p_{i, \vec{w}}^2 \rangle = \langle p_i^2 \rangle / 3$.
This allows to formulate the pressure as:
\begin{equation}\label{eq:Pressure}
 P = \sqrt{\langle p_\text{in}^2 \rangle / 3} \cdot \sqrt{\langle p_\text{out}^2 \rangle / 3} \cdot \frac{N}{m \cdot V} \;\text{,}
\end{equation}
with the volume $V = A \cdot L$.
In there, the averaged squared momentum $\langle p_i^2 \rangle$ of a macroscopic state $i$ is calculated as the integrated squared microcanonical momentum $p^2(E) = 2 \cdot m \cdot E$ weighted by the respective number of states $\rho(E) \cdot f_i(E)$, normalized by the integrated number of states:
\begin{equation}\label{eq:sqMomentum}
 \langle p_i^2 \rangle = \frac{\int\limits_0^\infty p^2(E) \cdot \rho(E) \cdot f_i (E) \cdot \text{d}E}{\int\limits_0^\infty \rho(E) \cdot f_i (E) \cdot \text{d}E} \text{,}
\end{equation}
with the density of states $\rho (E)$ and the energy distribution $f_i(E)$ of the macroscopic state $i$.
For a monoatomic particle, the isochoric density of states is obtained through inverse Laplace transformation of the translational partition function~\cite{Doentgen2016b} and is independent of the macroscopic state:
\begin{equation}\label{eq:DOS}
 \rho (E) = \frac{q_\text{tr} \cdot V}{\Gamma(3/2)} \cdot \sqrt{E} \text{,}
\end{equation}
with the translational partition function pre-factor $q_\text{tr} = (2 \pi m / h^2)^{3/2}$ and the gamma function $\Gamma$.
The energy distribution $f_i(E)$ describes the probability of encountering a particle with energy $E$.

\subsection{Impingement Rate}
The impingement rate $\langle j \rangle$ of a gas on a surface is closely related to the pressure, as it defines the number of collisions transferring the above averaged squared momentum to the surface.
This impingement rate is defined as the number of collisions with a container wall over time and surface area and is given as follows~\cite{Atkins2010pp753}:
\begin{equation}
 \langle j \rangle = \frac{N}{4V} \cdot \langle v \rangle \;\text{,}
\end{equation}
with the number of particles $N$, the volume $V$, and the average velocity $\langle v \rangle$.
When resolving the average velocity as the integrated microcanonical velocity $v = \sqrt{2 E / m}$, the average impingement rate is defined through the following integral.
\begin{equation}\label{eq:impingement}
 \langle j \rangle = \int\limits_0^\infty \underbrace{\frac{N}{4V} \cdot \sqrt{\frac{2 E}{m}}}_{:= j(E)} \cdot \frac{\rho(E) \cdot f_i(E)}{\int\limits_0^\infty \rho(E) \cdot f_i(E) \cdot \text{d}E} \cdot \text{d}E \;\text{,}
\end{equation}
with the microcanonical impingement rate $j(E)$.

\subsection{Heat Flux}
The averaged heat flux $\langle \dot{q} \rangle$ is obtained through integration of the microcanonical heat flux $\dot{q}(E)$ weighted with the number of states.
The microcanonical heat flux in turn is the microcanonical impingement rate times the energy transferred through collision $(E - E^\prime)$ weighted with the respective collisional energy transfer probability $P(E, E^\prime)$.

\begin{equation}\label{eq:heatflux}
 \resizebox{0.5\textwidth}{!}{$
  \begin{split}
  \langle \dot{q} \rangle = &\int\limits_0^\infty \underbrace{\frac{\int\limits_0^\infty j(E) \cdot (E - E^\prime) \cdot P(E, E^\prime) \text{d}E^\prime}{\int\limits_0^\infty P(E, E^\prime) \text{d}E^\prime}}_{:= \dot{q} (E)} \\
	&\cdot \frac{\rho(E) \cdot f_i(E)}{\int\limits_0^\infty \rho(E) \cdot f_i(E) \cdot \text{d}E} \cdot \text{d}E
	\end{split}
	$}
\end{equation}

For simplicity, it is assumed that the collisional energy transfer probability follows the single-exponential down model widely used for gas-gas collisional energy transfer~\cite{Carstensen2007}.

\begin{equation}\label{eq:colprop}
 P(E, E^\prime) = \begin{cases}
	A \cdot \exp\left(-\frac{E - E^\prime}{\langle \Delta E_\text{down} \rangle}\right), &E \geq E^\prime \\
	A \cdot \exp\left(-\frac{E^\prime - E}{\langle \Delta E_\text{up} \rangle}\right), &E < E^\prime
 \end{cases}
\end{equation}

In there, $\langle \Delta E_\text{down} \rangle$ is the average energy transferred downwards, $\langle \Delta E_\text{up} \rangle$ is the average energy transferred upwards, and $A$ is a normalizing factor.
In contrast to gas-gas collisions, gas-solid collisions do not necessarily have to fulfill the microscopic reversibility of collisional energy transfer with the gas phase~\cite{Carstensen2007}, since energy transferred to the solid can be assumed to diffuse away rather rapidly.
Here, the average upward energy transfer is modeled as $\langle \Delta E_\text{up} \rangle = \langle \Delta E_{\text{up}, 0} \rangle \cdot (T_\text{s}/T_0)^n$~\cite{Doentgen2019}, with the reference $\langle \Delta E_{\text{up, 0}} \rangle$ at reference temperature $T_0$, the temperature exponent $n$, and the surface temperature $T_\text{s}$.
The average downward energy transfer is then obtained by setting the average heat flux $\langle \dot{q} \rangle$ of a Boltzmann energy-distributed gas to zero for $T_\text{g} = T_\text{s}$, with the gas and surface temperatures, respectively.
Combining equations~\ref{eq:heatflux} and \ref{eq:colprop} yields the following expression for the microcanonical heat flux.

\begin{equation*}\label{eq:qE}
 \resizebox{0.5\textwidth}{!}{$
 \begin{split}
  \dot{q}(E) &= \frac{\int\limits_0^\infty j(E) \cdot (E - E^\prime) \cdot P(E, E^\prime) \text{d}E^\prime}{\int\limits_0^\infty P(E, E^\prime) \cdot \text{d}E^\prime} \\
  &= j(E)\frac{\int\limits_0^E (E - E^\prime) \cdot \exp\left(-\frac{E - E^\prime}{\langle \Delta E_\text{down} \rangle}\right) \cdot \text{d}E^\prime + \int\limits_E^\infty (E - E^\prime) \cdot \exp\left(-\frac{E^\prime - E}{\langle \Delta E_\text{up} \rangle}\right) \cdot \text{d}E^\prime}{\int\limits_0^E \exp\left(-\frac{E - E^\prime}{\langle \Delta E_\text{down} \rangle}\right) \text{d}E^\prime + \int\limits_E^\infty \exp\left(-\frac{E^\prime - E}{\langle \Delta E_\text{up} \rangle}\right) \text{d}E^\prime} \\
  &= j(E) \frac{\langle \Delta E_\text{down} \rangle^2 \cdot \left(1 - \frac{\langle \Delta E_\text{down} \rangle + E}{\langle \Delta E_\text{down} \rangle} \cdot \exp\left(-\frac{E}{\langle \Delta E_\text{down} \rangle}\right) \right) - \langle \Delta E_\text{up} \rangle^2}{\langle \Delta E_\text{down} \rangle \cdot \left( 1 - \exp\left(-\frac{E}{\langle \Delta E_\text{down} \rangle} \right) \right) + \langle \Delta E_\text{up} \rangle}
 \end{split}
 $}
\end{equation*}

\section{Validation}
For each of the three microcanonically formulated properties: Pressure, impingement rate, and heat flux, the present derivations are validated against well-established formulations for thermal equilibrium in the following.

\subsection{Pressure}
The present microcanonical formulation of pressure provided through equations~\ref{eq:Pressure}, \ref{eq:sqMomentum}, and \ref{eq:DOS} are validated for thermal equilibrium, i.e. for a Boltzmann energy distribution $f_\text{B}(E) = \exp(-E / k_\text{B} T)$, with the Boltzmann constant $k_\text{B}$ and the temperature $T$.
For this Boltzmann energy distribution, the averaged squared momentum is given as follows.
\begin{equation*}
 \resizebox{0.5\textwidth}{!}{$
 \begin{split}
  \langle p_\text{B}^2 \rangle &= \frac{\int\limits_0^\infty p(E)^2 \cdot \rho(E) \cdot f_B (E) \cdot \text{d}E}{\int\limits_0^\infty \rho(E) \cdot f_B (E) \cdot \text{d}E} \\
  &= \frac{\int\limits_0^\infty 2 m E \cdot \sqrt{E} \cdot \exp\left(-\frac{E}{k_\text{B} T}\right) \cdot \text{d}E}{\int\limits_0^\infty \sqrt{E} \cdot \exp\left(-\frac{E}{k_\text{B} T}\right) \cdot \text{d}E} \\
  &= \int\limits_0^\infty 2 m E \cdot \underbrace{2\sqrt{\frac{E}{\pi}} \cdot \left(\frac{1}{k_\text{B} T}\right)^{3/2} \cdot \exp\left(-\frac{E}{k_\text{B} T}\right) \cdot \text{d}E}_{\text{Maxwell-Boltzmann distribution}} \\
  &= 3 m k_\text{B} T
 \end{split}
 $}
\end{equation*}
Using the averaged squared momentum of a Boltzmann energy-distributed monoatomic gas in equation~\ref{eq:Pressure} yields the ideal gas law $P = N k_\text{B} T / V$.

In the present study, the delta energy distribution given by $f_\delta(E) = \delta(E - E_0)$ with the position of the delta peak $E_0$ is used as upper limiting case of a non-Boltzmann energy distribution.
To allow for comparability between different energy distributions, the total energy of the compared energy distributions is required to be equal, yielding $E_0 = 3/2 \cdot k_\text{B} T$.
With this expression for $E_0$ and equation~\ref{eq:sqMomentum}, the averaged squared momentum of a delta energy-distributed gas is given as follows.

\begin{equation*}
 \begin{split}
  \langle p_\delta^2 \rangle &= \frac{\int\limits_0^\infty p(E)^2 \cdot \rho (E) \cdot f_\delta(E) \cdot \text{d}E}{\int\limits_0^\infty \rho (E) \cdot f_\delta(E) \cdot \text{d}E} \\
  &= \frac{\int\limits_0^\infty 2 m E \cdot \sqrt{E} \cdot \delta(E - E_0) \cdot \text{d}E}{\int\limits_0^\infty \sqrt{E} \cdot \delta(E - E_0) \cdot \text{d}E} \\
  &= 2 m E_0 = 3 m k_\text{B} T
 \end{split}
\end{equation*}

Apparently, the averaged squared momenta of delta energy-distributed and Boltzmann energy-distributed monoatomic ideal gases are equal.

As proposed by D\"{o}ntgen~\cite{Doentgen2023}, the transition from the Boltzmann to the delta energy distribution is modeled through the Gaussian energy distribution $f_\text{G}(E) = \frac{1}{\sqrt{2 \pi} \cdot b} \cdot \exp\left(-\frac{(E - E_{\text{G},0})^2}{2 b^2}\right)$ with the Gaussian width $b$ and the Gaussian position $E_{\text{G},0}$.
When requiring that the total energy of the Gaussian and Boltzmann energy distributions are equal, $\langle E_\text{G} \rangle \overset{!}{=} \langle E_\text{B} \rangle$, the Gaussian width $b$ can be obtained numerically for a fixed Gaussian position $E_{\text{G},0}$~\cite{Doentgen2023}.
This allows to converge the Gaussian distribution either to the Boltzmann distribution for $E_{\text{G},0} \to \text{-}\infty$ or to the delta distribution for $E_{\text{G},0} \to 3/2 \cdot k_\text{B} T$.
When numerically evaluating equation~\ref{eq:sqMomentum} for the Gaussian distribution, one arrives at the same averaged squared momentum $\langle p_\text{G}^2 \rangle = 3 m k_\text{B} T$ as for the Boltzmann and delta distributions.

From this theoretical consideration, it can be concluded that the observable pressure is independent of the energy distribution of the monoatomic ideal gas exerting this pressure to the surface.
As a consequence, pressure measurements are not expected to carry information about the energy distribution of monoatomic ideal gases.

\subsection{Impingement Rate}
Evaluating equation~\ref{eq:impingement} yields the well-established impingement rate for Maxwell-Boltzmann distributed gases when assuming a Boltzmann energy distribution $f_\text{B}(E)$~\cite{Atkins2010pp753}:

\begin{equation*}
 \resizebox{0.5\textwidth}{!}{$
 \begin{split}
 \langle j_\text{B} \rangle &= \frac{N}{4V} \int\limits_0^\infty \sqrt{\frac{2 E}{m}} \cdot 2 \sqrt{\frac{E}{\pi}} \left ( \frac{1}{k_\text{B} T}\right )^{3/2} \cdot \exp\left(-\frac{E}{k_\text{B} T}\right) \cdot \text{d}E \\
 &= \frac{N}{4V} \cdot \sqrt{\frac{8 k_\text{B} T}{\pi m}} \text{.}
 \end{split}
 $}
\end{equation*}

For a delta energy distribution, equation~\ref{eq:impingement} and the above derived $E_0 = 3/2 \cdot k_\text{B} T$ provides the averaged impingement rate according to:

\begin{equation*}
 \begin{split}
 \langle j_\delta \rangle &= \frac{N}{4V} \frac{\int\limits_0^\infty \sqrt{\frac{2 E}{m}} \cdot \sqrt{E} \cdot \delta(E - E_0)\text{d}E}{\int\limits_0^\infty \sqrt{E} \cdot \delta(E - E_0)\text{d}E} \\
 &= \frac{N}{4V} \cdot \sqrt{\frac{2 E_0}{m}} \\
 &= \frac{N}{4V} \cdot \sqrt{\frac{3 k_\text{B} T}{m}} \text{.}
 \end{split}
\end{equation*}

The ratio of the Boltzmann impingement rate $\langle j_\text{B} \rangle $ and the delta impingement rate $\langle j_\delta \rangle$ provides the relative change of impingement when replacing a Boltzmann energy distribution with a delta energy distribution.

\begin{equation*}
 \frac{\langle j_\delta \rangle}{\langle j_\text{B} \rangle} = \sqrt{\frac{3 \pi}{8}} \approx 1.085
\end{equation*}

According to the impingement rate ratio, wall collisions are about 8.5\;\% more frequent in a delta energy-distributed gas compared to a Boltzmann energy-distributed gas.
Given the fact that the pressure, i.e. the momentum transfer, exerted through these wall collisions is independent of the energy distribution, a larger impingement rate refers to a smaller average momentum transfer per collision.
It is important to note that this statement refers to the averaged momentum transfer, not the averaged squared momentum transfer, which is unaffected by the energy distribution (cf. preceding section).

\begin{figure}[!htb]
 \centering
 \includegraphics[width=3in, keepaspectratio]{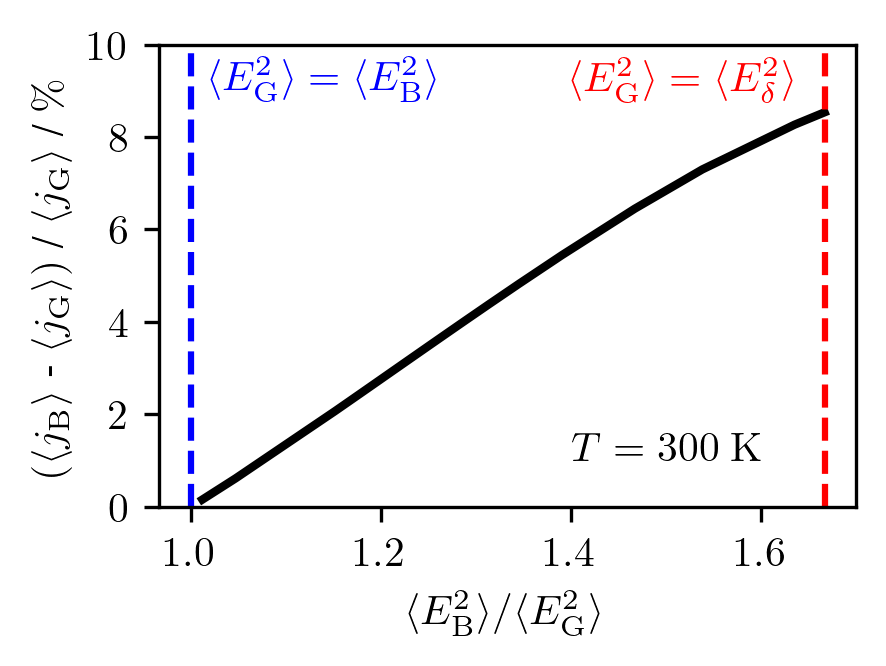}
 \caption{Ratio of impingement rates of Boltzmann and Gaussian distributed gases, $\langle j_\text{B} \rangle$ and $\langle j_\text{G} \rangle$, respectively, over the Gaussian position $E_{\text{G},0}$.}
 \label{fig:impingement}
\end{figure}

Again, the Gaussian distribution is used to interpolate between the Boltzmann and delta distributions.
Figure~\ref{fig:impingement} shows the ratio of impingement rates of a Boltzmann energy distributed gas $\langle j_\text{B} \rangle$ and a Gaussian energy distributed gas $\langle j_\text{G} \rangle$ as function of the ratio of averaged squared energies of the Boltzmann and Guassian distributions $\langle E_\text{B}^2 \rangle / \langle E_\text{G}^2 \rangle$.

The Gaussian impingement rate equals the delta impingement rate for $\langle E_\text{G}^2 \rangle \to \langle E_\delta^2 \rangle$ and converges to the Boltzmann impingement rate as $\langle E_\text{G}^2 \rangle \to \langle E_\text{B}^2 \rangle$.
As the Gaussian distribution approaches the delta distribution, the impingement rate ratio appears to further increase, yet a Gaussian distribution with $E_{\text{G},0} > E_0$ cannot satisfy $\langle E_\text{G} \rangle \overset{!}{=} \langle E_\text{B} \rangle$ anymore.
Therefore, the delta distribution stands as upper limiting case of the Gaussian distribution in the present study.

In contrast to the pressure, measuring the impingement rate is extremely difficult.
An interesting approach was developed by Arakawa and Tuzi~\cite{Arakawa1981} in 1981, who used calorimetric measurements of the energy transferred through condensation of gases on a cryogenic surface.
This surface was cooled to 7.5\;K through liquid helium and the impinging gas was at an initial temperature of 300\;K.
For Ar, Kr, Xe, \ce{CH4}, and \ce{N2}, experimentally determined impingement rates agreed with thermodynamically estimated ones with a 5\;\% systematic deviation.
For \ce{CO2}, however, the authors reported larger discrepancies and explained them with the more complex condensation mechanism of \ce{CO2} on cryogenic surfaces.
Arakawa and Tuzi~\cite{Arakawa1981} attributed the observed 5\;\% systematic error to uncertainties in the calorimeter calibration.
Interestingly, this deviation is comparable to the effect non-Boltzmann energy distributions can have on the impingement rate (cf. Figure~\ref{fig:impingement}).
Although this opens the possibility that Arakawa and Tuzi~\cite{Arakawa1981} actually measured the impingement rates of non-Boltzmann energy distributed gases, the uncertainties are too large to make a definite statement.

\subsection{Boltzmann Heat Flux from Theory}
\begin{figure}[!htb]
 \centering
 \includegraphics[width=3in, keepaspectratio]{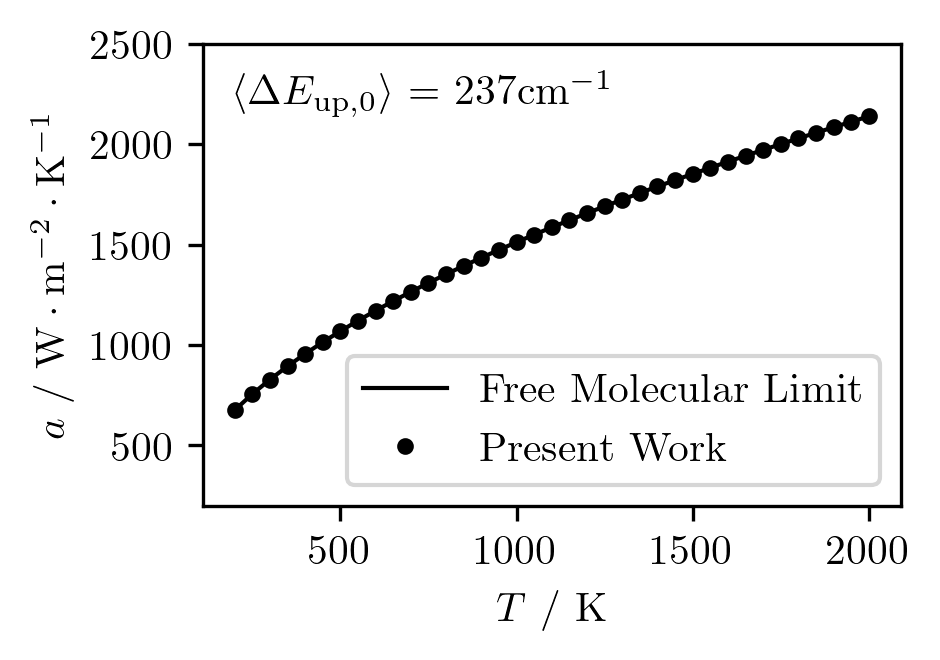}
 \caption{Comparison of the present heat transfer coefficient formulation (symbols) to the free molecular limiting case of the widely used Sherman-Lees equation (line)~\cite{Sherman1963, Lees1961, Bird1994, Trott2007}.}
 \label{fig:HeatFlux}
\end{figure}

In their experimental work, Arakawa and Tuzi~\cite{Arakawa1981} deduced impingement rates from the heat of condensation released by the adsorbing gas molecules.
In addition to the heat of condensation, the collisional energy transfer induces a heat flux into the surface or into the gas, depending on the surface and gas temperatures.
In order to validate the presented heat flux formulation in equation~\ref{eq:heatflux} for a Boltzmann energy-distributed monoatomic ideal gas, the heat transfer coefficient $a_\text{B} = \langle \dot{q}_\text{B} \rangle / (T_\text{g} - T_\text{s})$ with the above formulation for $\dot{q}(E)$ in equation~\ref{eq:qE} is compared to the free molecular (FM) limiting heat transfer coefficient, as described by the widely used Sherman-Lees equation~\cite{Sherman1963, Lees1961, Bird1994, Trott2007}.
\begin{equation}\label{eq:ShermanLees}
 a_\text{FM} = \frac{1}{2} \left( \frac{P \cdot \langle v \rangle}{T} \right) \left(\frac{\alpha}{2 - \alpha} \right) \left(1 + \frac{\zeta}{4}\right) \;\text{,}
\end{equation}
with the averaged velocity $\langle v \rangle$, the thermal accommodation coefficient $\alpha$, and the number of internal degrees of freedom of the gas $\zeta$.
This equation is simplified by assuming perfectly diffuse gas-solid collisions ($\alpha = 1$)~\cite{Trott2007} and a monoatomic gas ($\zeta = 0$) which obeys the ideal gas law ($P/T = k_\text{B} \cdot N/V$).
Moreover, the Sherman-Lees equation describes heat transfer between two opposing, parallel plates, thus includes both directions of the one-dimensional gas motion between these plates.
The present theoretical description, however, considers motion in only one direction, thus the free molecular limiting heat transfer coefficient defined in equation~\ref{eq:ShermanLees} has to be divided by $2$.
The resulting equation can be described in terms of the one-directional impingement rate $\langle j_\text{B} \rangle$ of a Boltzmann energy-distributed gas.
\begin{equation}
 a_\text{FM} = k_\text{B} \cdot \langle j_\text{B} \rangle
\end{equation}
In Figure~\ref{fig:HeatFlux}, this formulation of the free molecular limiting heat transfer coefficient is used as reference for the present formulation of the heat transfer coefficient provided through equations~\ref{eq:heatflux}, \ref{eq:colprop}, and \ref{eq:qE}.
Note that the reference upward energy transfer through gas-solid collisions $\langle \Delta E_{\text{up}, 0} \rangle = 237\;\text{cm}^{-1}$ has been used to fit the present formulation to the literature free molecular limiting heat transfer coefficient.
The literature formulation and the present formulation of the heat transfer coefficient match perfectly over the temperature range from 200 to 2000\;K.
Although the heat transfer coefficient obtained through the present theory was shifted to the free molecular limiting heat transfer coefficient by adjusting the upward reference collisional energy transfer $\langle \Delta E_{\text{up}, 0} \rangle$, the shape of the present heat transfer coefficient curve was not affected by this adjustment.
Therefore, it can be concluded that the presented theory correctly reproduces the literature formulation, with $\langle \Delta E_{\text{up}, 0} \rangle$ being a physical representation of the thermal accommodation process of gas-solid collisions.

\begin{figure}[!htb]
 \centering
 \includegraphics[width=3in, keepaspectratio]{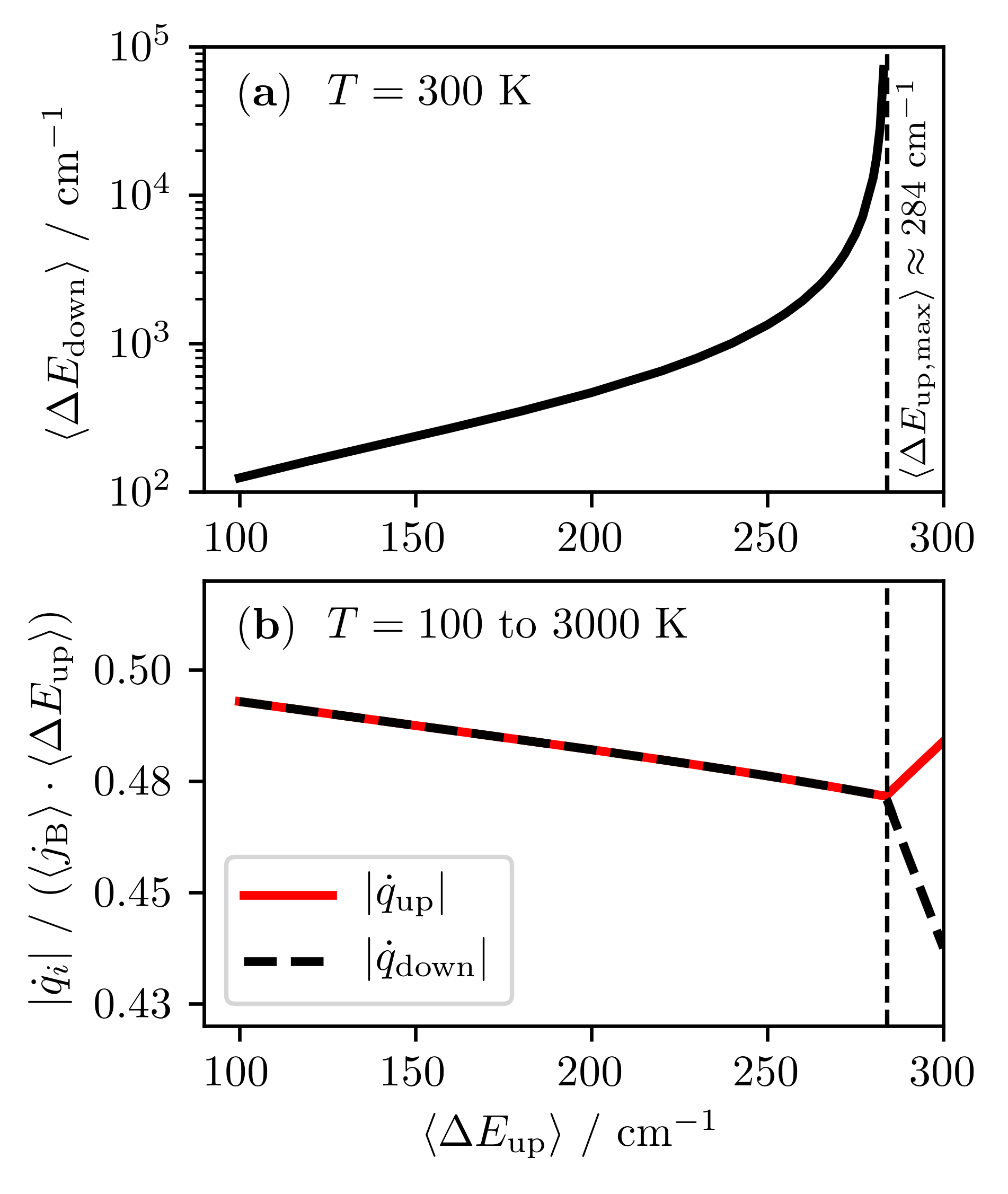}
 \caption{(a) Downward average energy transfer $\langle \Delta E_{\text{down}} \rangle$ as function of upward energy transfer $\langle \Delta E_{\text{up}} \rangle$ under the condition that $\lvert\dot{q}_\text{down}\rvert \overset{!}{=} \lvert\dot{q}_\text{up}\rvert$ for $T_\text{g} = T_\text{s}$.
 (b) Dimensionless upward and downward heat fluxes as function of $\langle \Delta E_{\text{up}} \rangle$.}
 \label{fig:dEdown}
\end{figure}

Interestingly, the presently used model for the collisional energy transfer probability (cf. equation~\ref{eq:colprop}) cannot describe heat transfer processes for $\langle \Delta E_{\text{up}, 0} \rangle \gtrsim 284\;\text{cm}^{-1}$.
Figure~\ref{fig:dEdown} shows how the downward average energy transfer $\langle \Delta E_{\text{down}} \rangle$ and the dimensionless upward and downward heat fluxes $ \lvert \dot{q}_i \rvert / (\langle j_\text{B} \rangle \cdot \langle \Delta E_\text{up} \rangle)$ depend on the upward average energy transfer $\langle \Delta E_{\text{up}} \rangle$ at $T = 300\;\text{K}$.
Note that the absolute value of the upward heat flux was used, so that it can be directly compared to the downward heat flux.
The total heat flux is separated into the downward and upward heat fluxes by separately integrating the first term (with $\langle \Delta E_{\text{down}} \rangle^2$) and the second term (with $-\langle \Delta E_{\text{up}} \rangle^2$) of $\dot{q}(E)$ (cf. equation~\ref{eq:qE}), respectively.
Further note that the downward average energy transfer $\langle \Delta E_{\text{down}} \rangle$ is calculated so that the net heat flux is zero for $T_\text{g} = T_\text{s}$, as described above.

For small values of $\langle \Delta E_{\text{up}} \rangle$, the downward average energy transfer $\langle \Delta E_{\text{down}} \rangle$ is proportional to $\langle \Delta E_{\text{up}} \rangle$, yet $\langle \Delta E_{\text{down}} \rangle$ is always larger than $\langle \Delta E_{\text{up}} \rangle$.
This can simply be explained by the integration of the collisional energy transfer, in which the upward energy transfer is always fully integrated from $E$ to $\infty$, while the downward energy transfer is cut off by the lower integration limit $0$ (cf. equation~\ref{eq:qE}).
Due to this truncated integration of the downward energy transfer, the respective average energy transfer $\langle \Delta E_\text{down} \rangle$ has to be larger than the upward average energy transfer $\langle \Delta E_{\text{up}} \rangle$ to compensate for the truncation.
Up to the observed $\langle \Delta E_{\text{up}} \rangle \approx 284\;\text{cm}^{-1}$ limit, Figure~\ref{fig:dEdown}(b) shows that the absolute downward and upward heat fluxes are equal, as imposed by the $\langle \Delta E_{\text{down}} \rangle$ calculation scheme.
When approaching the $\langle \Delta E_{\text{up}} \rangle$ limit, however, compensation of the truncated integration of the downward energy transfer leads to a diverging $\langle \Delta E_{\text{down}} \rangle$ (cf. Figure~\ref{fig:dEdown}(a)).
Beyond the $\langle \Delta E_{\text{up}} \rangle$ limit, $\langle \Delta E_{\text{down}} \rangle$ cannot compensate the truncated integration anymore and the present model is not valid.
For the temperature range from 100 to 3000\;K, this effect is temperature independent, since the dimensionless downward and upward heat fluxes do not depend on temperature.

\section{Methodology}
Molecular dynamics simulations of a nanostructure in contact with Boltzmann and non-Boltzmann energy-distributed monoatomic gas has been carried out to test the present non-Boltzmann heat flux theory.
Simulations have been carried out using the Lennard-Jones (LJ) interaction potential with the LAMMPS software package~\cite{Plimpton1995}.

Initially, a circular area in the center of the 450\;{\AA} wide and 900\;{\AA} high two-dimensional simulation box was filled with 341 solid phase atoms with a 6.16\;{\AA} lattice spacing.
The LJ parameters used to describe this solid have been selected based on the properties of elementary iron (Fe), with a molar mass of $M_\text{Fe} = 55.85\;\text{g}/\text{mol}$~\cite{LANL_Periodic_iron}, a LJ well depth of $\varepsilon_\text{Fe} = 23.9\;\text{kcal}/\text{mol}$, and a LJ diameter of $\sigma_\text{Fe} = 3.88\;\text{\AA}$.
The well depth is taken from the bond dissociation energy of iron-iron bonds~\cite{Dean1998} and the diameter is the doubled covalent radius of iron~\cite{LANL_Periodic_iron}.
The solid phase atoms have been energy-minimized, resulting in the formation of three grain-like sub-sections of the solid, which persist during the entire simulation.
This is due to the hexagonal minimum energy crystal structure of the Lennard-Jones atoms, which disrupts the initial circular edge of the solid phase (cf. Figure~\ref{fig:SimulationBox}).
The diameter of the minimum energy solid structure ranges from about 79\;{\AA} to 90\;{\AA}, thus the solid structure can be categorized as nanomaterial~\cite{Adamo2017}.
Before adding the gas phase atoms, the solid structure has been thermalized at the target solid temperatures $T_\text{s}$ for 0.5\;ns using a Nos\'{e}-Hoover thermostat~\cite{Nose1984, Hoover1985} with a damping factor of 10\;fs and a time step of 0.5\;fs.
Note that the center-of-mass velocity of the solid structure is forced to be zero during all simulations.

The area surrounding the solid structure has been filled with 698 gas phase atoms, the LJ parameters of which have been taken from elementary argon (Ar), with a molar mass of $M_\text{Ar} = 39.95\;\text{g}/\text{mol}$~\cite{LANL_Periodic_argon}, a LJ well depth of $\varepsilon_\text{Ar} = 114\;\text{K}$~\cite{Hippler1983} and a LJ diameter of $\sigma_\text{Ar} = 3.47\;\text{\AA}$~\cite{Hippler1983}.
For the present two-dimensional simulations the effective diameter of the gas-gas interactions was calculated so that the 2D and 3D hard sphere collision frequencies are equal~\cite{Doentgen2023} according to $\sigma_\text{eff.} = 2 \cdot \sigma^2 / L = 0.107\;${\AA}, with $L = 225\;\text{\AA}$ being the reference depth of the reduced z-direction in the present simulations.
The resulting effective gas density amounts to 12.72\;mol/$\text{m}^3$.
The effective diameter concept has been validated for the simulation of an ideal shock tube process previously~\cite{Doentgen2023}.
For the gas-solid and solid-solid interactions, the original LJ diameters have been used.
Therefore, the present simulation setup loosely resembles an infinitely expanded iron nanowire in an argon atmosphere at effective pressures of about 0.74 to 1.38\;bar for gas temperatures $T_\text{g}$ of 700 to 1300\;K.

For each gas temperature $T_\text{g}$ of 700, 900, 1100, and 1300\;K, solid temperatures with $T_\text{g} - T_\text{s}$ of 0, 50, 100, 150, 200, 250, and 300\;K have been simulated for 5\;ns with a time step of 0.5\;fs within the constant number of particles ($N$), constant volume ($V$), and constant energy ($E$) ensemble.
For each condition, 20 replica simulations with different random atomic positions and velocities have been carried out to reduce statistical uncertainties.
The initial atomic velocity distribution of the gas phase was either resembling a Maxwell-Boltzmann velocity distribution (Boltzmann case) or each atom at the exact same initial velocity (non-Boltzmann case according to the delta distribution).
The total energies of all gas particles have been evaluated every 500\;fs, averaged over all replica simulations, and traced over the simulated time to obtain the slope of the energy change, i.e. the heat flux between the gas and solid phases.
To roughly estimate the relevance of gas-gas collisions in the present simulations, the hard sphere collision frequency is used to calculate the statistically expected number of gas-gas collisions during the 5\;ns simulations, which amount to 6 -- 8 collision events in a single simulation.
Therefore, the presented non-Boltzmann heat flux results are practically unaffected by gas-gas interactions.

In addition to the above simulations without maintaining the delta distribution, single simulations for each gas temperature have been carried out for 50\;ns with the velocity rescaling routine for maintaining a delta energy-distributed gas~\cite{Doentgen2023}.
With these simulations, the equilibria between delta energy-distributed gases and the solid are evaluated.
In these simulations, the solid temperature was initially equal to the respective gas temperature.
The solid temperatures converged after the first 30\;ns of the simulations and the final solid temperatures were evaluated for the last 10\;ns of the simulations.

\section{Results and Discussion}
\subsection{(Non-)Boltzmann Heat Flux from Molecular Dynamics}

\begin{figure}[!htb]
 \centering
 \includegraphics[width=3in, keepaspectratio]{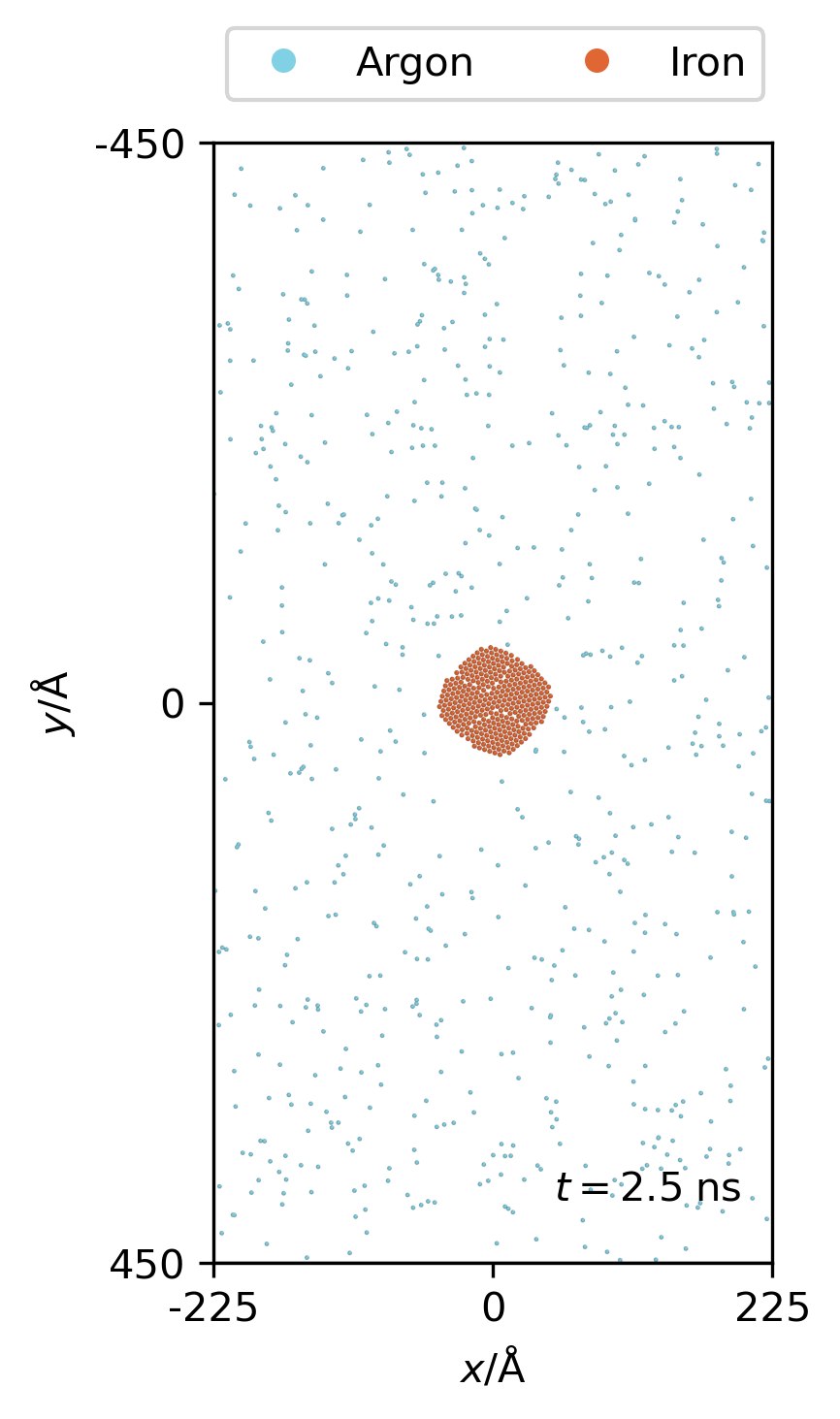}
 \caption{Exemplary snapshot of the present heat flux simulations}
 \label{fig:SimulationBox}
\end{figure}

Figure~\ref{fig:SimulationBox} illustrates the presently used simulation setup for determination of Boltzmann and non-Boltzmann heat fluxes for the above-described iron nanostructure in a low-pressure argon atmosphere.

The solid structure in the center shows two nearly symmetric grain boundaries between a central elongated grain and two smaller grains on each side of the solid structure.
The surface shape of the solid structure is loosely circular, but is slightly shifted towards a hexagonal shape, resulting in a smallest diameter of 79\;{\AA} and a largest diameter of 90\;{\AA}, as reported before.
The gas phase atoms are randomly distributed around the solid structure.

\begin{figure*}[!htb]
 \centering
 \includegraphics[width=6in, keepaspectratio]{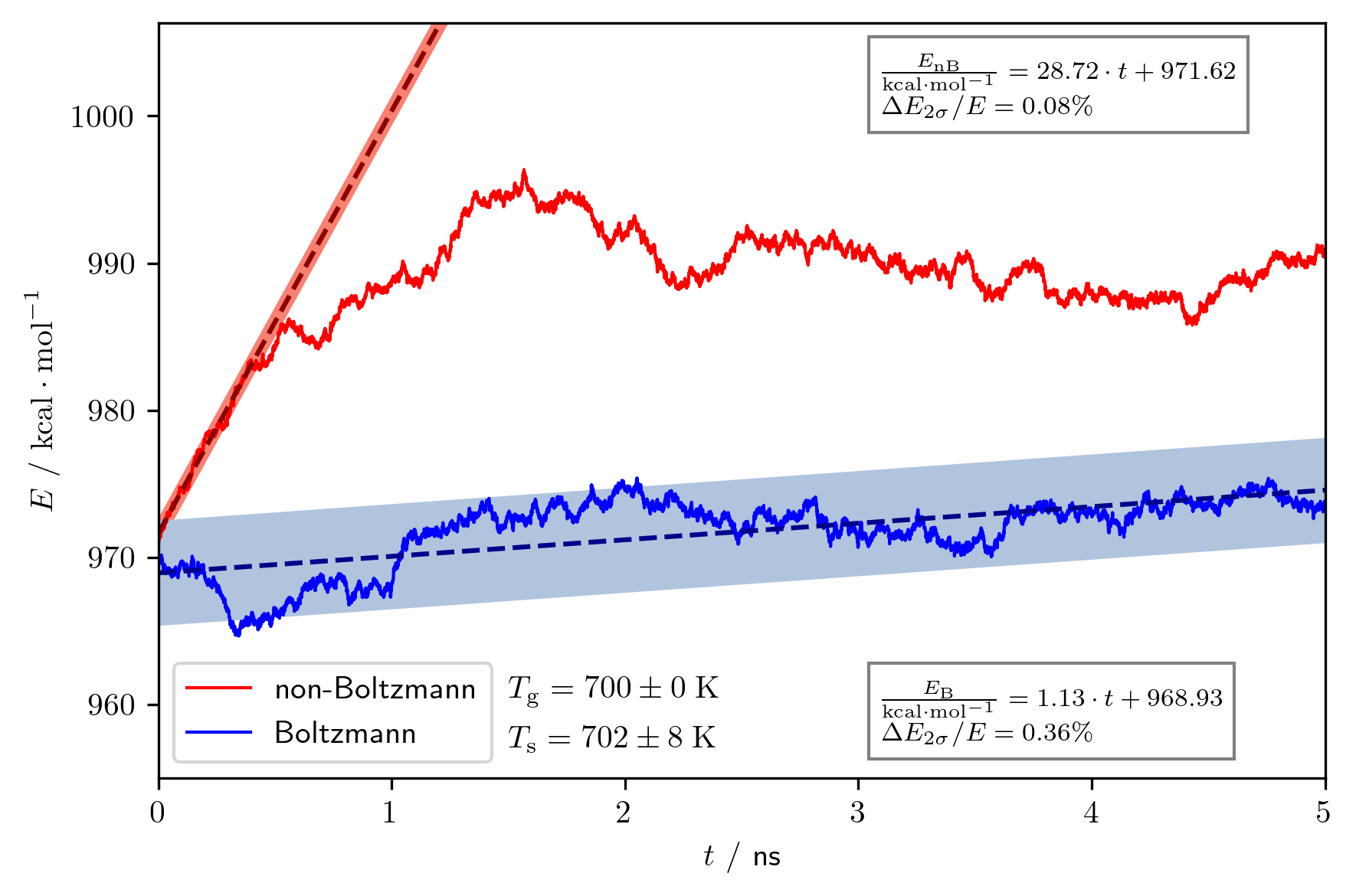}
 \caption{Exemplary energy profiles over simulation time for the Boltzmann and non-Boltzmann simulations.
The dashed lines represent linear fits to the initial energy changes.}
 \label{fig:EnergyProfiles}
\end{figure*}

Based on the trajectory simulations, the total energy of the gas phase atoms has been traced for the Boltzmann and non-Boltzmann cases and the slopes of the initial energy change in these profiles have been fitted linearly.
Figure~\ref{fig:EnergyProfiles} shows exemplary energy profiles over the simulation time of the Boltzmann and non-Boltzmann simulations for a gas and solid temperature of $T_\text{g} = T_\text{s} = 700\;\text{K}$.

In the Boltzmann case, the energy remains nearly constant on average and the observed slope of the linear fit is due to the slightly larger solid temperature compared to the gas temperature.
The energy fluctuations are due to the collisional energy transfer between the gas and solid phases and the solid phase essentially acts as heat bath for the gas phase.
Similar fluctuations are observed in molecular dynamics simulations with an actual or virtual heat bath, i.e. thermostat~\cite{Doentgen2015}.
In the non-Boltzmann case, the energy initially increases rapidly and appears to fluctuate at a higher energy compared to the Boltzmann case.
The initial change of energy observed in the non-Boltzmann simulation is clearly stronger than the energy fluctuations observed in the Boltzmann simulation.
It is concluded that the non-Boltzmann energy-distributed gas receives energy from the solid phase.
This process ceases once the energy distribution of the gas nearly converged to a Boltzmann energy distribution through collisions with the solid surface.

The slopes of the linear fits presented in Figure~\ref{fig:EnergyProfiles} are defined as heat fluxes here and the same fitting procedure has been used to evaluate all present simulations.
The resulting heat fluxes of all simulations are listed in Table~\ref{tab:HeatFluxes}.
Note that the gas and solid temperatures are the actual initial temperatures and that the solid temperature in particular has been fluctuating in the preceding thermalization, leading to a certain scatter in the initial solid temperatures.
Further note that some simulations did not succeed due to numerical issues, resulting in 18 to 20 replica for each condition, as listed in Tabel~\ref{tab:HeatFluxes}.

\begin{table*}
 \centering
 \caption{Boltzmann (B) and non-Boltzmann (nB) heat fluxes obtained via linear fitting of the initial energy change of the merged Boltzmann and non-Boltzmann replica simulations.}
 \label{tab:HeatFluxes}
 \begin{tabular}{lllll}
  $T_\text{g}$ & $T_\text{s}$ & $N_\text{rep}$ & $\dot{q}_\text{B}$ & $\dot{q}_\text{nB}$ \\
  K & K & - & $\frac{\text{kcal}}{\text{mol}\cdot\text{ns}}$ & $\frac{\text{kcal}}{\text{mol}\cdot\text{ns}}$ \\
	\hline
	\begin{filecontents*}{HeatFluxData.csv}
Tg,Tssim,dTs,Nrep,dEBdt,dEBE,dEnBdt,dEnBE
700,403,4,20,-19.8,0.94,-7.98,0.09
700,446,5,20,-13.42,0.79,-6.29,0.07
700,502,6,20,-13.82,0.56,2.19,0.09
700,544,6,20,-12.98,0.61,13.51,0.09
700,591,7,20,-6.74,0.48,18.46,0.12
700,652,8,20,-2.11,0.45,18.42,0.12
700,702,8,20,1.13,0.36,28.72,0.08
900,591,7,20,-27.23,1.19,5.63,0.09
900,652,8,20,-21.53,0.88,13.13,0.09
900,702,8,20,-20.37,0.63,4.25,0.09
900,748,9,20,-13.07,0.49,28.13,0.1
900,805,10,19,-5.65,0.37,47.64,0.13
900,836,10,20,-2.84,0.8,48.05,0.11
900,899,11,19,-2.84,0.64,48.59,0.11
1100,804,10,19,-27.95,1.01,18.71,0.09
1100,834,10,19,-20.67,1.6,16.41,0.12
1100,901,11,20,-9.98,1.67,57.6,0.12
1100,946,11,19,-11.42,1.18,47.56,0.11
1100,993,12,20,-4.72,0.9,69.93,0.11
1100,1050,13,19,4.31,0.58,67.06,0.07
1100,1104,14,19,1.86,0.53,87.99,0.12
1300,993,12,18,-17.78,1.37,40.74,0.13
1300,1052,12,20,-3.56,1.32,72.86,0.24
1300,1103,14,19,-8.11,1.28,63.67,0.2
1300,1152,14,20,-11.77,0.7,77.88,0.09
1300,1188,14,19,1.56,1.23,98.12,0.16
1300,1266,15,19,4.65,0.5,118.89,0.11
1300,1313,16,20,8.55,0.72,123.15,0.14
\end{filecontents*}
\csvreader[head to column names]{HeatFluxData.csv}{}%
	{\\ \Tg & \Tssim $\pm$ \dTs & \Nrep & \dEBdt $\pm$ \dEBE & \dEnBdt $\pm$ \dEnBE}%
	\\ \hline
 \end{tabular}
\end{table*}

\begin{figure*}[!htb]
 \centering
 \includegraphics[width=6in, keepaspectratio]{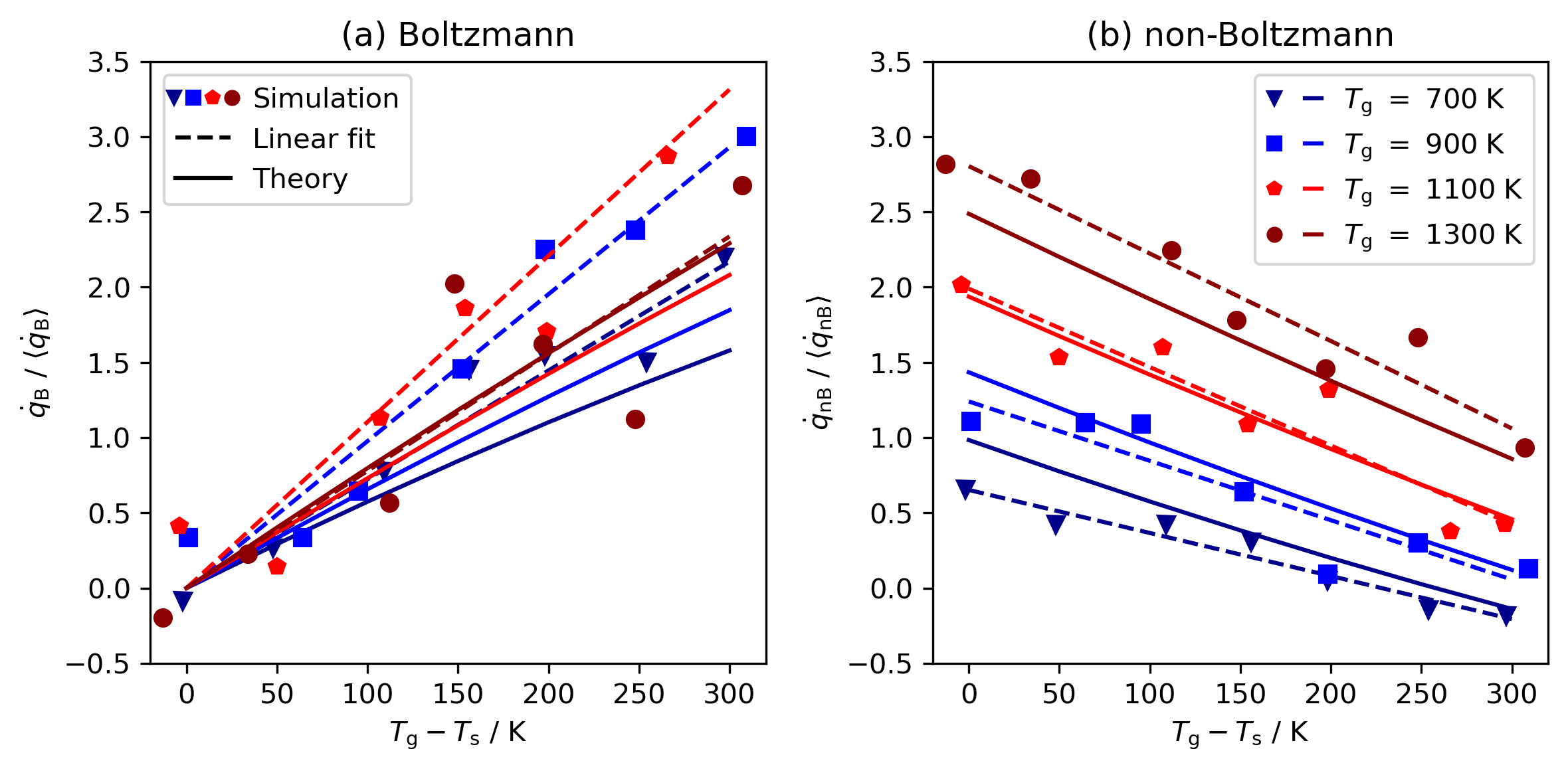}
 \caption{Relative heat fluxes for the Boltzmann (a) and non-Boltzmann (b) cases obtained via simulation and theory.
The dashed lines represent linear fits to the simulation results.}
 \label{fig:RelativeHeatFluxes}
\end{figure*}

Comparison of the Boltzmann and non-Boltzmann heat fluxes from the simulations to the theoretical predictions via equation~\ref{eq:heatflux} is done relative to the respective averaged heat fluxes to shift all results to the same scale.
Figure~\ref{fig:RelativeHeatFluxes} shows the present theoretical and simulation results for the Boltzmann\;(a) and non-Boltzmann\;(b) cases.
The dashed lines represent linear fits to the simulation results and allow for direct comparison to the solid lines, which have been obtained via theory.
For the Boltzmann case, any systematic offset for $T_\text{g} - T_\text{s} = 0$ was subtracted to allow for a fair comparison to the theoretical results, which are inherently yielding $\dot{q}_\text{B} = 0$ for this condition.

In the Boltzmann case (cf. \ref{fig:RelativeHeatFluxes}\;(a)), the expected trend of increasing heat flux from the hot gas to the cold solid is observed for increasing temperature difference $T_\text{g} - T_\text{s}$.
The gas temperature itself has a minor effect on the relative heat flux, but one can clearly observe an increasing relative heat flux with increasing gas temperature for the theoretical results.
Also the simulation results reflect this trend, except for linear fit to the highest temperature, the slope of which is unexpectedly small, which is likely due to statistical uncertainties of the simulations.
Although the theoretically predicted relative heat fluxes appear to come with smaller slopes compared to those of the linear fits to the simulation results, the scattering of the latter indicate that the differences in slopes might not be statistically significant.

In the non-Boltzmann case (cf. \ref{fig:RelativeHeatFluxes}\;(b)), both the simulation results and the theoretical predictions indicate that heat is flowing from the solid to the gas in case of $T_\text{g} - T_\text{s} = 0$, which is in grave contrast to the behavior observed in the Boltzmann case.
This effect is weakening with increasing temperature difference between the two phases, yet only for the lowest presently investigated gas temperature of $T_\text{g} = 700\;\text{K}$ the heat flux reaches zero if the gas is about 250\;K hotter than the cold solid.
For all other investigated gas temperatures, the hotter non-Boltzmann gas is always receiving energy from the colder solid.
Moreover, the impact of the gas temperature is significantly stronger compared to the Boltzmann case and higher gas temperatures appear to promote the non-Boltzmann heat flux from the cold solid to the hot gas.
When comparing the simulation results to the theoretical predictions, the qualitative trends are well matched, yet the gas temperature dependency is underestimated by the presently proposed theory.
This is likely due to the collisional energy transfer modeling for gas-solid collisions, which has been adopted from gas-gas collisions in the present study.



\begin{figure}[!htb]
 \centering
 \includegraphics[width=2.5in, keepaspectratio]{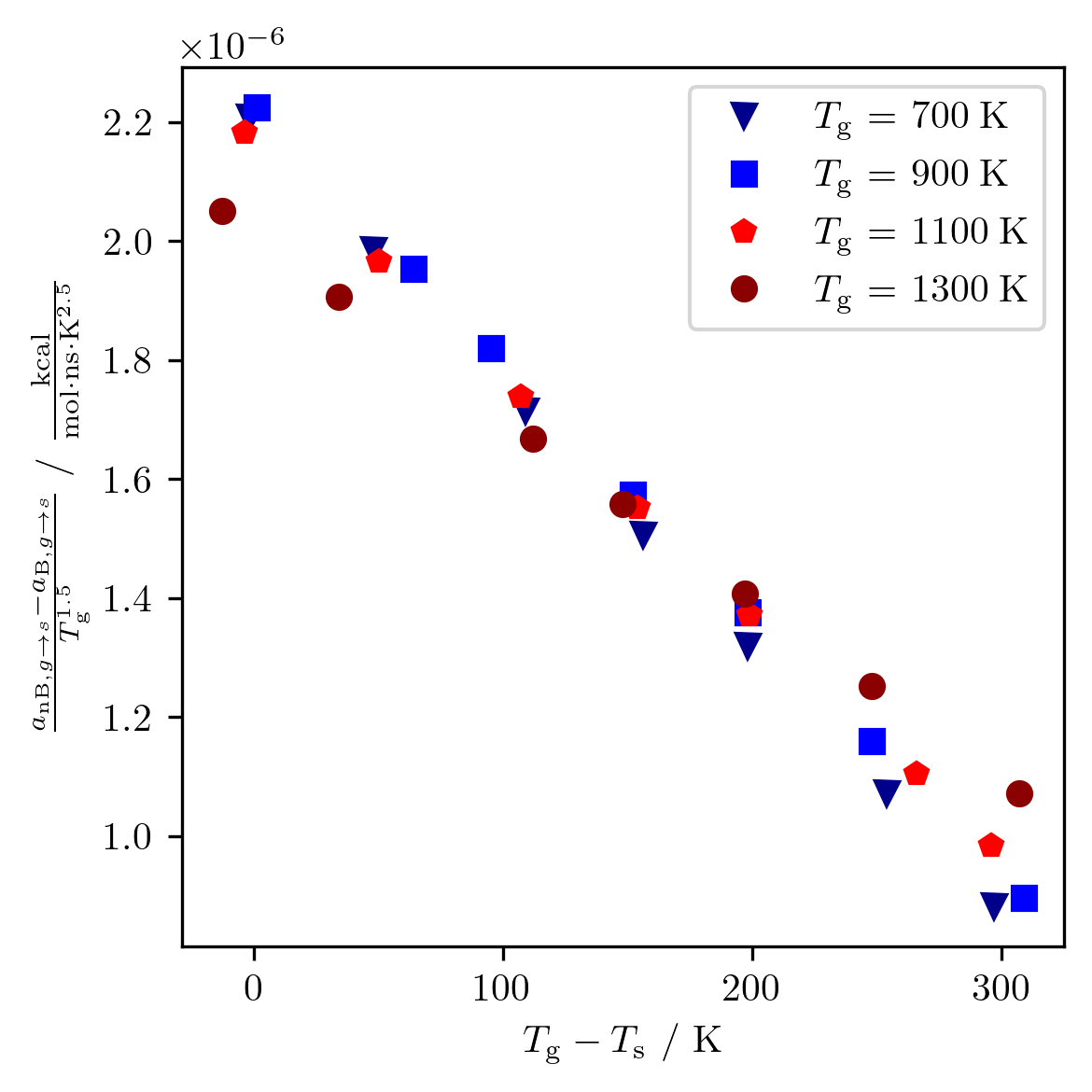}
 \caption{Difference of non-Boltzmann (nB) and Boltzmann (B) heat transfer coefficients $a$ relative to a power of the gas temperature.}
 \label{fig:Tdep}
\end{figure}

In an attempt to gain insights into the functional dependency of the non-Boltzmann heat flux, the difference between the non-Boltzmann and Boltzmann heat fluxes has been related to the gas temperature in Figure~\ref{fig:Tdep}.
In there, the heat transfer coefficients $a_\text{B}$ and $a_\text{nB}$ are used to represent the Boltzmann and non-Boltzmann cases, respectively.
The exponent of the gas temperature in the denominator was selected so that all curves collapse on a single line.

Two things are apparent in Figure~\ref{fig:Tdep}: Firstly, the exponent of $T_\text{g}$ amounts to 1.5, which is different from the expected 0.5 from the Sherman-Lees theory for the Boltzmann case (cf. equation~\ref{eq:ShermanLees}; Secondly, the gas temperature alone cannot explain the difference of the non-Boltzmann to the Boltzmann heat fluxes, as the temperature difference $T_\text{g} - T_\text{s}$ also affects this difference.
When reconsidering equation~\ref{eq:heatflux}, one can conclude that the temperature exponent must result from the non-Boltzmann weighting of the density of states in the second part of the equation, as the microcanonical heat flux $\dot{q}(E)$ does not depend on the energy distribution.
The stronger gas temperature-dependency of the non-Boltzmann heat flux can also be seen in Figure~\ref{fig:RelativeHeatFluxes}.
From a simplified point of view, the heat flux from the solid to the gas should cancel out when considering the difference between the non-Boltzmann and Boltzmann heat fluxes.
Apparently, this is not true and either the heat flux from the solid to the gas is different in the non-Boltzmann and Boltzmann cases or the temperature of the solid couples into the heat flux from the gas to the solid.
Future research will resolve this ambiguity.

\subsection{Delta Distribution Equilibrium State}
The observed non-Boltzmann heat flux effect, i.e. the transfer of thermal energy from a cold solid to a hot gas, has been further tested by converging a single simulation for each gas temperature to its equilibrium state, while maintaining the delta energy distribution.
In each of these simulations, the average solid temperature converged to a steady value after no more than 30\;ns.
The final solid temperatures have been calculated by averaging over the last 10\;ns of the 50\;ns simulations, meaning that the solid temperatures did not change beyond the expected fluctuations during the averaging period.
Table~\ref{tab:DeltaEq} lists the nominal gas temperatures of the delta energy-distributed gas phase and the corresponding equilibrium solid temperatures.

\begin{table}
 \centering
 \caption{Equilibrium solid temperatures for delta energy-distributed gas.}
 \label{tab:DeltaEq}
 \begin{tabular}{ll}
  $T_\text{g}$ & $T_\text{s}$ \\
  K & K \\
	\hline
	700 & 403$\pm$16 \\
	900 & 503$\pm$21 \\
	1100 & 586$\pm$25 \\
	1300 & 674$\pm$29 \\
	\hline
	\hline
 \end{tabular}
\end{table}

The present equilibrium solid temperatures appear to be proportional to the gas temperature with a factor of 0.54, meaning that a solid phase in equilibrium with a delta energy-distributed gas phase is 46\;\% colder than the gas phase.
Note that this factor has been obtained by fitting the data in Table~\ref{tab:DeltaEq} with a y-axis intersection at 0\;K, as the gas cannot receive energy from the solid at 0\;K by definition.
Despite being of pure theoretical interest due to the short-lived nature of the delta energy distribution, the observed temperature difference is remarkable.
Future studies will be investigating the equilibrium solid temperatures of Gaussian energy distributions in between the limiting cases defined through the Boltzmann and delta energy distributions.

\section{Conclusions}
The present work provides microcanonical formulations for the pressure, impingement rate, and heat flux of a monoatomic gas with an arbitrary energy distribution in contact with an ideal solid surface.
These formulations have been validated against well-established models for the case of a Boltzmann energy-distributed gas.
Testing the validated formulations for non-Boltzmann energy distributions revealed that pressure is not affected by the energy distribution.
The impingement rate, however, is found to be larger by up to 8.5\;\% for the non-Boltzmann case compared to the Boltzmann case.
The heat flux, in particular, is severely affected by deviations from the Boltzmann energy distribution and can even become negative, meaning that heat flows from a cold solid to a hot gas.

The theoretically proposed behavior of the non-Boltzmann heat flux was tested via molecular dynamics simulations of an iron nanowire in an argon atmosphere at low pressures.
Although the temperature of the solid nanostructure was never larger than that of the gas in these simulations, the gas gained energy through gas-solid collisions.
The present simulations and the present theory both indicate that heat flows from the cold solid to the hot gas and show the same qualitative trend.
This strongly indicates that the presently proposed non-Boltzmann heat flux effect is physically meaningful.
Equilibrium simulations for the upper limiting non-Boltzmann distribution, the delta distribution, indicate that the solid phase would be 46\;\% colder than the gas at equilibrium, if this energy distribution could be maintained.

The theoretically proposed and computationally observed non-Boltzmann heat flux effect has been reported for the first time, to the best of the authors' knowledge.
Therefore, no experimental proof of the presently proposed behavior is available as of today.
Future work will focus on designing an experiment which will be capable of proving the non-Boltzmann heat flux effect.
Moreover, it is necessary to advance theory and simulation from monoatomic to diatomic and polyatomic gases, so that it can be tested against existing experiments which involve non-equilibrium energy distributions.

The present work provides novel insights into the behavior of non-Boltzmann gases and reveals the counter-intuitive non-Boltzmann heat flux effect, which leads to heat transfer from cold solids to hot gases.
Since heat transfer is key for describing the thermodynamics of many natural and technical systems, the present findings might help to better understand known thermodynamical processes and could potentially allow developing novel processes.



\bibliography{literature}

\end{document}